\documentclass{jaa}
\usepackage{natbib}
\def\aap {{A\&A\,}}

\def\apj {{ApJ\,}}
\def\apjs {{ApJS\,}}
\def\apjl {{ApJL\,}}
\def\aj {{AJ\,}}

\def\araa  {{ARA\&A\,}}
\def\mnras {{MNRAS\,}}

\def\lbol {$L\textsubscript{bol}~$}
\def\tbol {$T\textsubscript{bol}~$}

\newcommand*{\mydprime}{$^{\prime\prime}\mkern-1.2mu$}
\usepackage{graphicx}

\begin{document}\sloppy

\title{Optical spectroscopy of Gaia detected protostars with DOT: can we probe protostellar photospheres? }


\author{Mayank Narang\textsuperscript{1,4}, Manoj. P\textsuperscript{1}, Himanshu Tyagi\textsuperscript{1}, 	Prasanta K. Nayak\textsuperscript{1,5}, Saurabh  Sharma\textsuperscript{2}, Arun Surya\textsuperscript{1}, Bihan  Banerjee\textsuperscript{1},  Blesson Mathew\textsuperscript{3}, Arpan Ghosh\textsuperscript{2}, Aayushi Verma\textsuperscript{2}  }
\affilOne{\textsuperscript{1}Department of Astronomy and Astrophysics, Tata Institute of Fundamental Research (TIFR) 
Homi Bhabha Road, Colaba, Mumbai 400005, India. \\}
\affilTwo{\textsuperscript{2} Aryabhatta Research Institute of Observational Sciences (ARIES), Manora Peak, Nainital 263 001, India \\} \affilThree{\textsuperscript{3} Department of Physics and Electronics, CHRIST (Deemed to be University), Bangalore 560029, India \\}  \affilFour{\textsuperscript{4} Academia Sinica Institute of Astronomy \& Astrophysics, 11F of Astro-Math Bldg., No.1, Sec. 4, Roosevelt Rd., Taipei 10617, Taiwan, R.O.C. \\}
\affilFive{\textsuperscript{5} Institute of Astrophysics, Pontificia Universidad Catolica de Chile, Av. Vicuña MacKenna 4860, 7820436, Santiago, Chile\\} 


\twocolumn[{

\maketitle

\corres{mnarang@asiaa.sinica.edu.tw}


\begin{abstract}
Optical spectroscopy offers the most direct view of the stellar properties and the accretion indicators. Standard accretion tracers, such as $H\beta$, $H\alpha$, and, Ca II triplet lines, and most photospheric features, fall in the optical wavelengths. However, these tracers are not readily observable from deeply embedded protostars because of the large line of sight extinction (Av $\sim$ 50-100 mag) toward them. In some cases, however, it is possible to observe protostars at optical wavelengths if the outflow cavity is aligned along the line-of-sight that allows observations of the photosphere, or the envelope is very tenuous and thin such that the extinction is low. In such cases, we can not only detect these protostars at optical wavelengths but also follow up spectroscopically. We have used the HOPS catalog (Furlan et al. 2016) of protostars in Orion to search for optical counterparts for protostars in the Gaia DR3 survey. Out of the 330 protostars in the HOPS sample,  an optical counterpart within 2" is detected for 62 of the protostars. For 17 out of 62 optically detected protostars, we obtained optical spectra  { (between 5500 to 8900 $\AA$) using the Aries-Devasthal Faint Object Spectrograph \& Camera (ADFOSC) on the 3.6-m Devasthal Optical Telescope (DOT) and Hanle Faint Object Spectrograph Camera (HFOSC) on 2-m Himalayan Chandra Telescope (HCT)}. We detect strong photospheric features, such as the TiO bands in the spectra {(of 4 protostars)}, hinting that photospheres can form early on in the star formation process. We further determined the spectral types of protostars, which show photospheres similar to a late M-type. Mass accretion rates derived for the protostars are similar to those found for T-Tauri stars, in the range of 10$^{-7}$ to 10$^{-8}$ $M_\odot$/yr. 
\end{abstract}

}]


\doinum{12.3456/s78910-011-012-3}
\artcitid{\#\#\#\#}
\volnum{000}
\year{0000}
\pgrange{1--}
\setcounter{page}{1}
\lp{1}

\section{Introduction}

One of the central questions in star formation is, when does a star attain its final mass? The relatively low mass accretion rates observed for T-Tauri stars ($\sim 10^{-8} M_\odot/yr $) \citep[e.g.,][]{Hartmann98} suggest that the star must acquire most of its mass during the protostellar phase \citep[e.g.,][]{white04}. It is also in the protostellar phase that the protoplanetary disk is formed, and the initial physical and chemical conditions for planet formation are set \citep[e.g.,][]{Vorobyov, Kratter10, Li14, Tobin20,  2023ApJS..266...32P}. {Measuring the accretion rate and accretion feedback from protostars is essential in comprehending the process of star and planet formation. However, determining the accretion rates of protostars poses a challenge. The conventional accretion tracers like H$\beta$, H$\alpha$, and Ca-II IR triplet lines \citep{Hartmann16} are situated in the optical wavelengths and are not easily observable due to significant line-of-sight extinction (Av $\sim$ 50 mag) \citep{Whitney97} towards deeply embedded protostars. Protostars have been extensively studied in the near-infrared (NIR) range, where accretion tracers like Pa$\beta$ and Br$\gamma$ can be utilized to quantify the mass accretion rates \citep[e.g.,][]{1997AJ....114.2157G,D05,2018ApJ...862...85G,E22}. However, all the NIR signatures of accretion are secondary tracers, and the UBV continuum excess \citep{Hartmann16}, which is the direct tracer of accretion, lies in the optical wavelengths.  }

{Optical detection and observations of protostars have been few and far between \citep[e.g.,][]{Kenyon98, white04, Riaz15}.} Optical spectroscopy has played a central role in advancing our understanding of low-mass star formation, especially pre-main sequence evolution. It is the most accurate tool for characterizing the stellar photospheric properties (e.g., $T_{\mathrm{eff}}$ or spectral type, $\mathrm{log~\textsl{g}}$, and [Fe/H]) \citep{white04}. With optical spectra of embedded protostars, we can address the long-standing question of ``When does the photosphere first form in protostars''? By detecting the photospheric features, we can assign a spectral type to the protostars, which has long been challenging. pectra at wavelengths $\leq$~1~$\mu$m can provide a direct estimate of the mass accretion rate, an important probe on the mass accretion/ejection processes in protostars, and constrain the physical conditions of accretion shock models \citep{white07}. {Thus, measuring stellar photospheric properties and probing the processes at the star-disk interface requires spectroscopic observations at the optical wavelengths, where protostars are barely visible.}

{Most studies on protostar are carried out in the NIR or longer wavelength regimes where these sources are relatively bright \citep[e.g.,][]{1997AJ....114.2157G, D05, 2018ApJ...862...85G, 2022ApJ...937L...9F, E22}. However, there are two significant advantages of optical spectroscopy compared to NIR spectroscopy when it comes to characterizing stellar and accretion properties \citep{white04}:}

\begin{enumerate}
    \item {Optical spectroscopy provides a more direct view of the stellar properties and the accretion luminosity. In the NIR range, the emission can be dominated by thermal emission from warm dust and gas. However, in the optical range, the light is primarily emitted from the photosphere and the accretion shocks. This distinction allows optical spectroscopy to offer a clearer and more direct understanding of the stellar properties and the energy released during accretion.}

    \item {Optical wavelengths are more efficient at scattering by small dust grains compared to NIR wavelengths. Consequently, even when the stellar photosphere is not directly visible due to high circumstellar extinction, the presence of cavities can enable the observation of the photosphere and accretion shocks through scattered light. This scattering phenomenon provides an opportunity to gather valuable information about the stellar and accretion processes that would otherwise be obscured in the NIR spectrum.} 
    
\end{enumerate}

Thus, measuring stellar photospheric properties and probing the processes at the star-disk interface requires spectroscopic observations at the optical wavelengths, where protostars are barely visible.

{The dust in the protostar envelope reprocesses the shorter-wavelength photons emitted by the central star, and the accretion shocks and re-emits at mid-to far-infrared wavelengths \citep{2010ApJ...710..470D,Dunham14}. Due to this, the SEDs of the protostars peak in the far-IR, and little optical emission is detected from protostars. However, in some cases, it is possible to observe the protostar at optical wavelengths provided the envelope cavity of the protostar is aligned favorably with respect to the line-of-sight so as to permit the observations of the photosphere and accretion shock through scattered light \citep{white07, Kenyon98} or the envelope is very tenuous and thin such that the extinction towards the protostar is low (Narang et al., in prep). In such cases, it is possible even to obtain optical spectra of the protostars \citep[e.g.,][]{Kenyon98, white04, Riaz15}. }

In this work, we leverage the sensitivity and spatial resolution of $Gaia$ to detect faint objects, even in crowded regions. Since the protostars are deeply embedded, $Gaia$ provides us with an extensive and deep survey that can be used to detect some of them. We present a sample of 62 protostars from the Orion star-forming region with an optical counterpart detected in the $Gaia$ EDR3/DR3 \citep{Gaiae3}. Section 2 discusses the sample and the $Gaia$ cross-matching process. In Section 3, we investigate the differences between the protostellar properties of the parent sample and the optically detected sample: we compare the bolometric luminosity and bolometric temperature. Section 4 presents the optical spectra of some of the optically bright protostars; we estimate the spectral types and derive the mass accretion rates onto these protostars. Finally, we summarize our findings in Section 5.

\begin{figure*}
\centering
\includegraphics[width=0.8\linewidth]{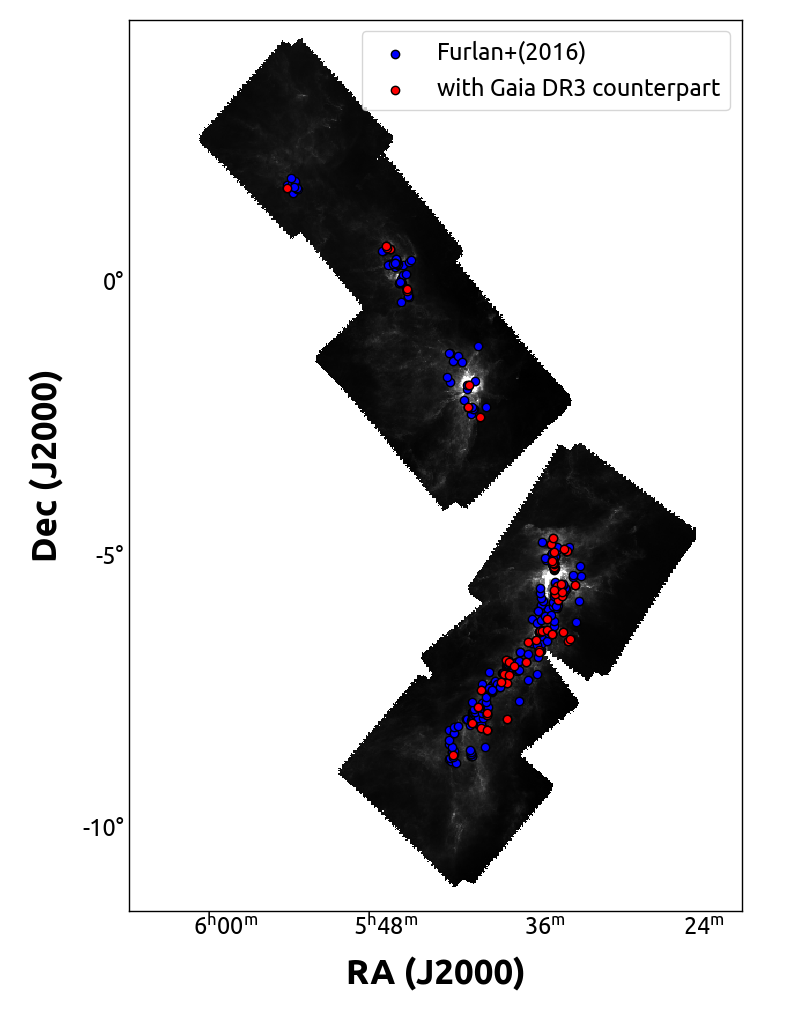}
\caption{The spatial distribution of the HOPS sources in the OMC region overlaid on the SPIRE 250 $\mu$m image. The HOPS sources with optical counterparts are shown as solid red circles, and the HOPS sources from the parent sample from \cite{Furlan16} are shown as solid blue circles.    }
\label{ch2fig1}
\end{figure*}
 
\section{Sample Selection}
To ensure a homogeneous sample of protostars with well-determined properties and comprehensive follow-up observations, we used the protostellar sample from Herschel Orion Protostellar Survey (HOPS) {\citep{Tom, Manoj13, Furlan16,2023ApJ...944...49F}}. The HOPS  sample consists of 92 Class 0 protostars, 125 Class I protostars, 102 flat-spectrum sources, and 11 Class II sources totaling 330 objects. This is the largest sample of protostars studied from a single star formation region within 500 pc from us. For these protostars, photometry data from 2MASS, \textit{ Spitzer}, \textit{Herschel}, APEX, and ALMA have been compiled, and the protostellar properties are robustly and homogeneously determined.

To cross-match the HOPS sources with $Gaia$ EDR3/DR3 \citep{Gaiae3}, we used the source position listed in \cite{Furlan16} and cross-matched it with the $Gaia$ EDR3/DR3 catalog. We used a search radius of 2\mydprime, which resulted in 68 cross-matches for 62 HOPS sources. In addition, six of the HOPS sources have another companion detected with $Gaia$ DR3  within 2\mydprime. These sources are HOPS 45, HOPS 71, HOPS 163, HOPS 170, HOPS 293, and HOPS 378. This work analyzes the optical source closest to the IR position. 

The final sample of $Gaia$ detected HOPS protostars consisted of 62 sources with 18 Class I sources, 36 flat-spectrum sources, and 8 Class II sources. No Class 0 protostars were detected with Gaia. All the $Gaia$ detected sources are within 1.3\mydprime~ of the  IR position listed in \cite{Furlan16} with a median offset of only 0.3\mydprime. So far, this is the largest sample of optically detected protostars from a single star-forming region. This work has increased the number of optically visible protostars currently known by a factor of 2-3.   

\begin{figure}
\centering
\includegraphics[width=1\linewidth]{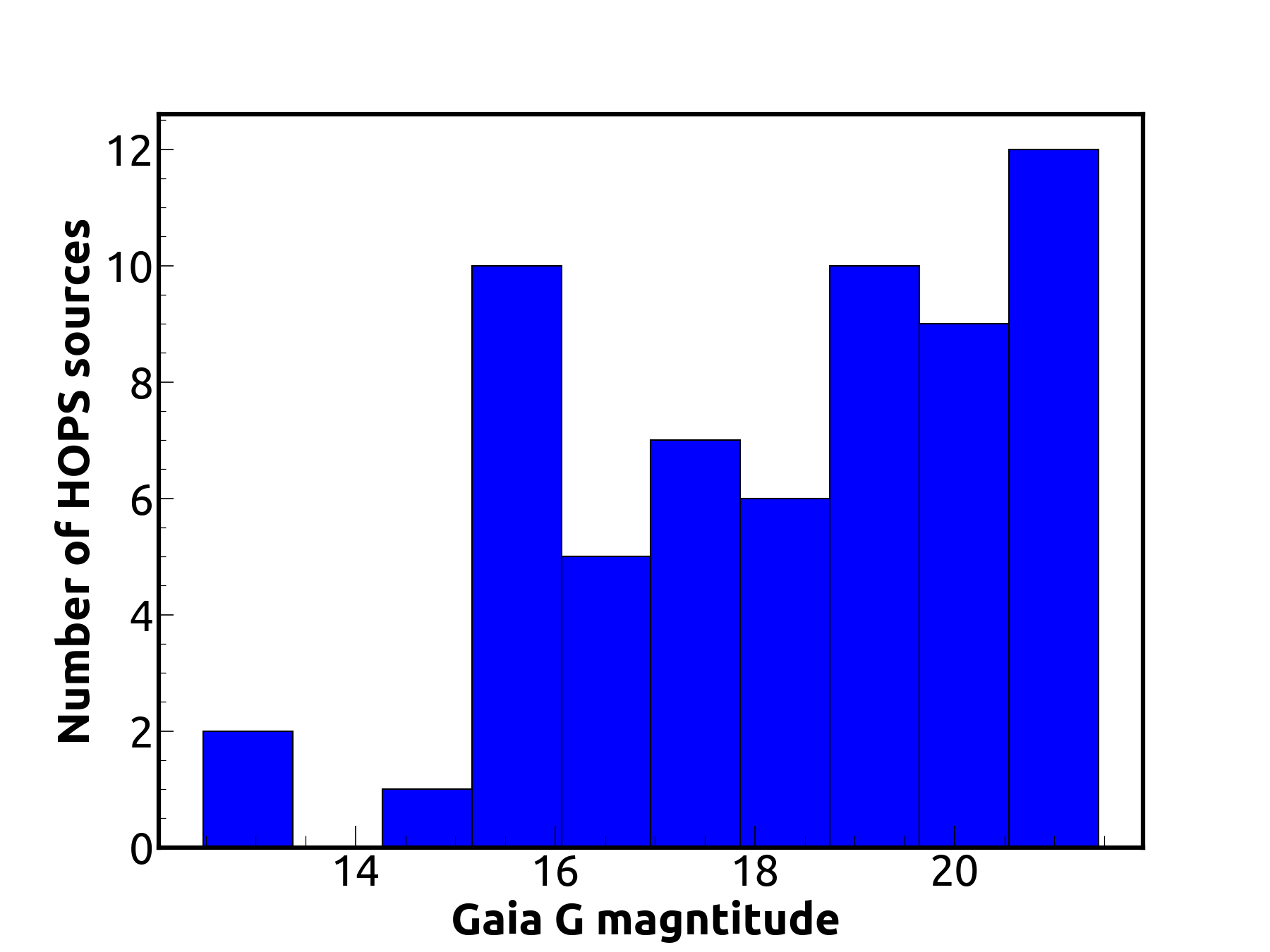}
\caption{The $Gaia$ G band magnitude of the optically detected HOPS sources.}
\label{ch2fig3}
\end{figure}

{The spatial distribution of the parent sample and the HOPS sources with $Gaia$ DR3 counterparts are shown in Figure \ref{ch2fig1}. There is no preference for the $Gaia$ detected sources to be either predominantly from  Orion A or Orion B. Since the protostars have a large line of sight extinction, they are extremely faint at shorter wavelengths. This is the major reason why not many studies on the optical properties of protostars have been carried out.} In Figure \ref{ch2fig3}, we show the distribution of $Gaia$ G band magnitude for the optically detected HOPS sources. The optically detected sources are faint, with a median magnitude of 18 in the G band.

\section{Evolutionary status}
{Among the 62 sources detected optically, 60\% exhibit flat-spectrum characteristics, while roughly 30\% belong to Class I sources. Interestingly, no Class 0 sources have been detected at optical wavelengths using Gaia. The initial sample from which these detections were made consisted of 27\% Class 0 sources, 40\% Class I sources, 30\% flat-spectrum sources, and 3\% Class II sources. Comparing the optically detected sources to the parent sample, it appears that the optically detected sources are more evolved. }

\begin{figure*}
\centering
\includegraphics[width=0.5\linewidth]{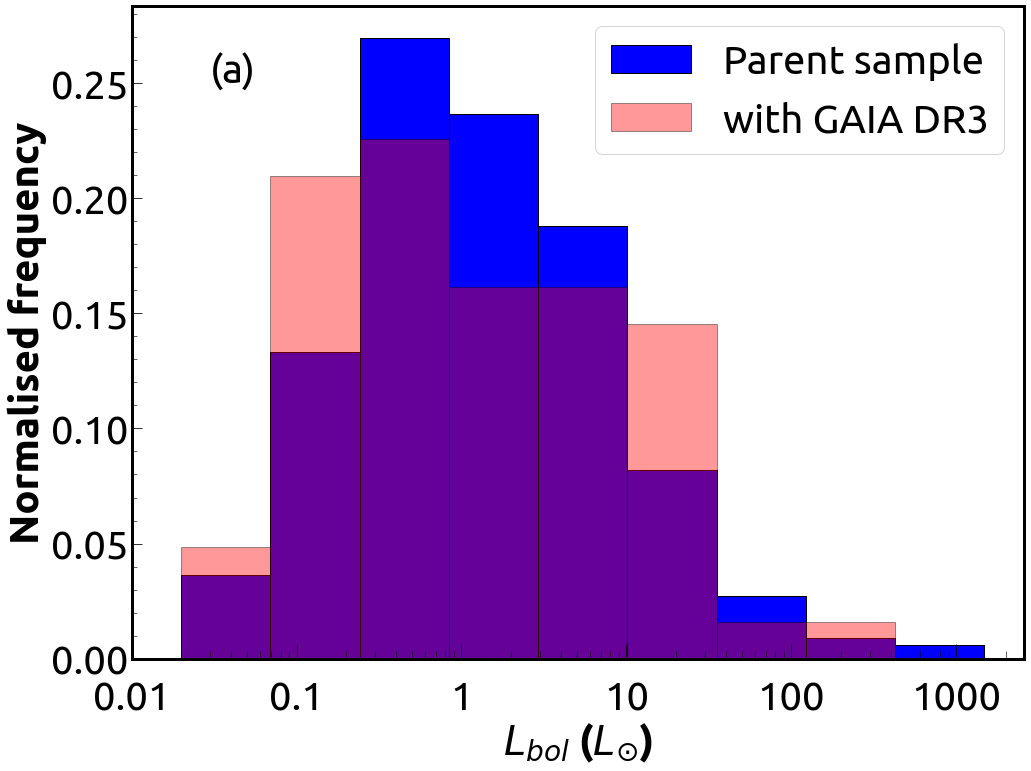}\includegraphics[width=0.49\linewidth]{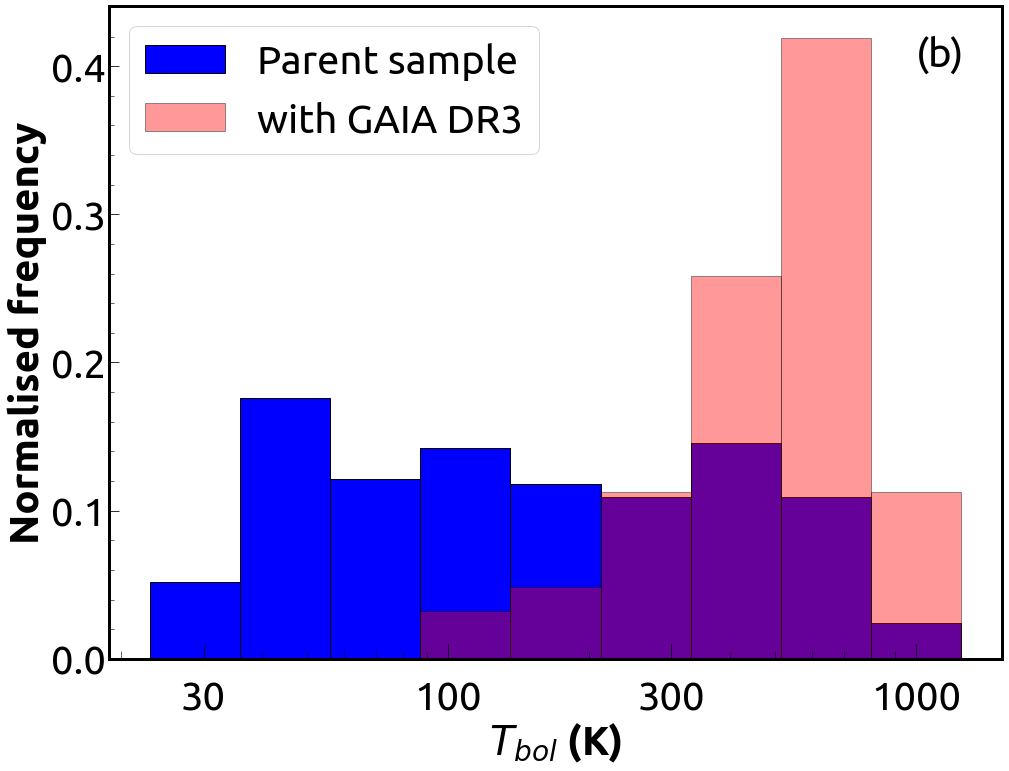}
\caption{(a) The \lbol~ distribution of the parent sample (blue histogram) and the HOPS sources with Gaia DR3 counterparts (peach histogram). {The Y axis represents the fraction of the HOPS target in each sample (either optically detected or the larger parent sample).} The \lbol~ distribution is similar between the two samples. (b)  The \tbol distribution of the parent sample (blue histogram) and the HOPS sources with Gaia DR3 counterparts (peach histogram).   }
\label{ch2fig5}
\end{figure*}

We compare the bolometric luminosity, \lbol and bolometric temperature, \tbol of the parent sample, and the optically detected sources. {We use the \lbol and \tbol listed in \cite{Furlan16} for this purpose.} In Figure \ref{ch2fig5} (a) we have shown the \lbol distribution. To test whether the \lbol distribution of the parent sample and the sources with optical counterparts are similar or different, we use the two-sample two-sided Kolmogorov-Smirnov test \citep{2020NatMe..17..261V}.

The $L_\mathrm{bol}$ distribution of the parent sample and the sources with optical counterparts have a two-sample KS statistic D = 0.11 and an associated p-value of $47\%$. Since the D value is small and the p-value is large, we cannot reject the null hypothesis in favor of the alternative. This suggests that the \lbol distribution of the parent sample and optically detected protostars are similar and likely drawn from the same parent sample.

The \tbol~ distribution (see Figure \ref{ch2fig5} (b)) of the optically detected sample is, on average higher than the parent sample. The two-sample KS  test on the \tbol~ values of the two samples gives us the two-sample KS statistic D = 0.55 and an associated p-value of $2 \times 10^{-13}\%$. The low p-value indicates that the probability that the two \tbol~ distributions are drawn from the initial sample distribution is vanishingly small, suggesting that the \tbol distributions of these two samples are statistically different.

The higher \tbol values of the optically detected HOPS sources indicate that they are more evolved than the parent sample. This is also consistent with the fact that most of the optically detected protostars are flat-spectrum sources and Class~I sources, and we do not detect any Class~0 sources. Since these sources appear to be more evolved than the parent sample, they have likely dissipated some of their envelope material, making detecting them at the optical wavelengths easier.

\section{Optical spectroscopy of the HOPS targets}

As shown in Figure \ref{ch2fig3}, the median $Gaia$ G band magnitude of the optically detected protostars is $\sim$ 18. Some protostars are bright enough to be observed using 2-4 m class telescopes. For these HOPS sources, we obtained low-resolution (R $\sim$ $1100-2500$) optical spectra. We observed these protostars using the  Aries-Devasthal Faint Object Spectrograph $\&$ Camera (ADFOSC) on the 3.6-m Devasthal Optical Telescope (DOT) and the  Hanle Faint Object Spectrograph Camera  (HFOSC) on the Himalayan Chandra Telescope (HCT). The spectra were obtained between December 2019 to January 2022. We observed 16 HOPS sources with the ADFOSC instrument (as part of DOT-C1-P50-2021; PI-Mayank Narang) using the 676R Grism (3500-8950 \AA) with a slitwidth of 1\mydprime. We also observed optical spectra of 6 protostars using HFOSC using Grism8 (5000-9100 \AA) and a slit width of 1.92\mydprime. { We obtained spectra for  flat-spectrum 17 protostars, with multi-epoch spectra for three protostars.}

We used the Hfosc-Automated-PIpeLIne (HAPILI)\footnote{https://github.com/Mayankattifr/HAPILI} (also see Appendix) to reduce the HFOSC data. HAPILI is an automated pipeline to reduce data of the Faint Object Spectrograph (FOS) class of instruments and was developed for the Hanle Faint Object Spectrograph Camera (HFOSC) on HCT. We further modified the HAPILI pipeline to reduce ADFOSC data as well. HAPILI is a python-based pipeline that runs PYRAF  as the back-end. The pipeline performs dark, bias correction, and flat fielding and automatically does the aperture identification and extraction. This is followed by wavelength calibration to produce the final spectra. 

\subsection{Detection of photospheric features in protostars}

 {Several broadband (TiO $\lambda\lambda$ 6250, 6800, 7140;  VO $\lambda\lambda$ 7510 \citealt{white04, Herczeg14}) and narrow absorption (e.g., Ca I $\lambda\lambda$  6156.023, 6471.662; Fe I $\lambda\lambda$ 5295.312, 6151.617, 6469.193; K I $\lambda\lambda$7665, 7699; Na I $\lambda\lambda$ 8183, 8195;  \citealt{white04,2018A&A...610A..40R}) features are present in the 5000-9100 \AA~range in the spectra of main-sequence and T-Tauri stars}. These features have been extensively used for spectral typing T-Tauri stars \citep[e.g.,][]{Luh04, Luh06, Herczeg14}. Some of these features were also used by \cite{white04} spectral-type protostars detected at optical wavelengths. For 4 out of the 17 protostars, the photospheric feature of TiO {7140} \AA~band is detected in the spectra, suggesting that the photosphere can form early on in the star formation process. In these cases, the central (proto)stars appear to have a photosphere similar to that of a late M-type star. 

\begin{figure*}
\centering
\includegraphics[scale=0.27]{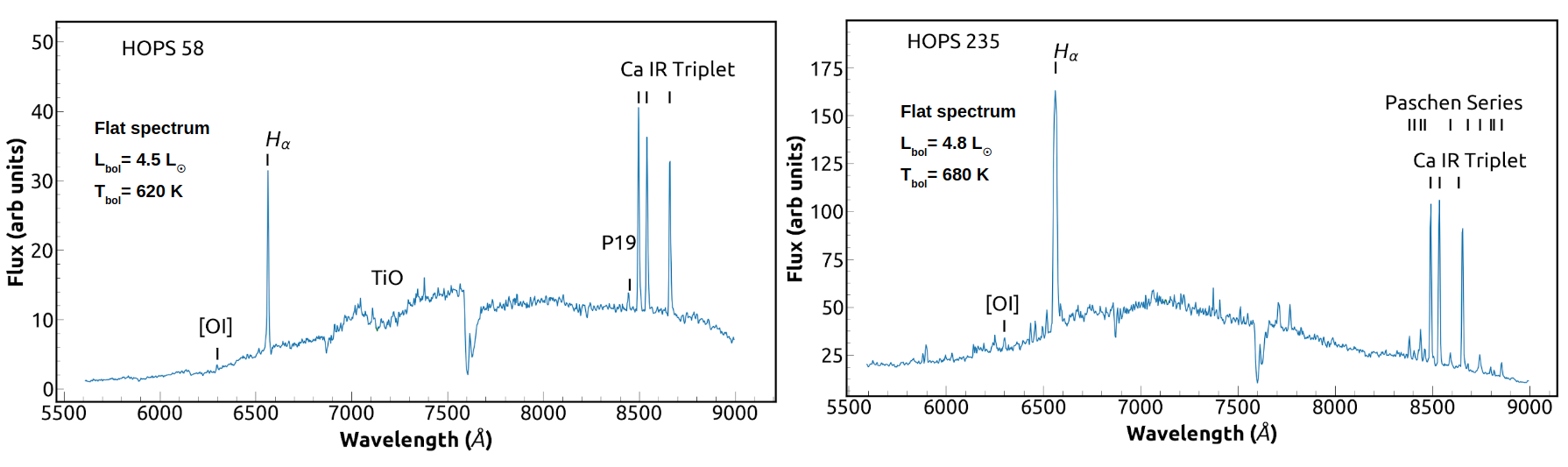}
\caption{The optical spectra taken with ADFOSC on DOT for HOPS 58  and HOPS 235  were obtained as part of {DOT-2021-C1-P50 (PI-Mayank Narang)}. Also identified are the prominent lines and TiO band (the spectra, however, is not flux calibrated). These two protostars have very similar protostellar properties but have different spectral profiles. }
\label{ch2fig1aa}
\end{figure*}

In Figure~\ref{ch2fig1aa}, we show the ADFOSC spectra of HOPS 58 and HOPS 235 obtained as part of DOT-2021-C1-P50 (PI-Mayank Narang). Both HOPS 58 and HOPS 235 have similar $L_\mathrm{bol}$ and $T_\mathrm{bol}$, yet HOPS 58 displays the TiO feature, while HOPS 235 does not show the feature, indicating that even for protostars having similar protostellar properties ($L_\mathrm{bol}$ and $T_\mathrm{bol}$), the optical spectra can be very different. This dichotomy in the spectra of essentially similar protostars can be due to two possible reasons: either the photosphere is yet to form in HOPS 235, or that HOPS 235 has a hotter photosphere (spectral type earlier than M-type) and hence does not have the TiO feature.

Even though we detected the broadband TiO feature, no narrowband features were detected in the spectra. There are two main reasons why we haven't detected any narrow photospheric lines: (i) accretion shocks can produce an underlying $\sim$ 10,000 K blackbody continuum, which can fill in the absorption lines leading to their diminished strength. This is known as veiling \citep{Veil1, Veil2}. If the veiling is strong, as might be the case for protostars, the entire line can be filled in, resulting in the absence of the narrow features in the spectrum. (ii) Another possible reason for the non-detection of narrow photospheric features could be the low SNR of our spectra. Nonetheless, from this analysis, the main conclusion that we can draw is that the photospheres can form fairly early, {at least in some cases, as early as the flat-spectrum stage. }

\subsection{Mass accretion rates from protostars}

We computed mass accretion rates for the protostars for which we have optical spectra. We use the relation between accretion and line luminosity derived for T-Tauri stars. A similar prescription has been used by \cite{Riaz15}, who used relations derived for T-Tauri stars to compute the mass accretion rates from protostars. {From the optical spectra, we measured the equivalent widths of $H\alpha$; then estimated the continuum underlying the $H\alpha$ line from $Gaia$ $G_{RP}$ \citep[similar to][]{2018ApJ...857...30M} that is dereddened using the Av values listed in \cite{Furlan16} following the prescription in \cite{Arun}. We computed the $H\alpha$ luminosity ($L_{H \alpha}$) using these quantities. In the case of T-Tauri stars, it has been shown that $L_{H \alpha}$ is correlated with accretion luminosity ($L_{\mathrm{acc}}$) \citep[e.g.,][]{Muz,Alcala} such that }

\begin{figure}
\centering
\includegraphics[scale=0.2]{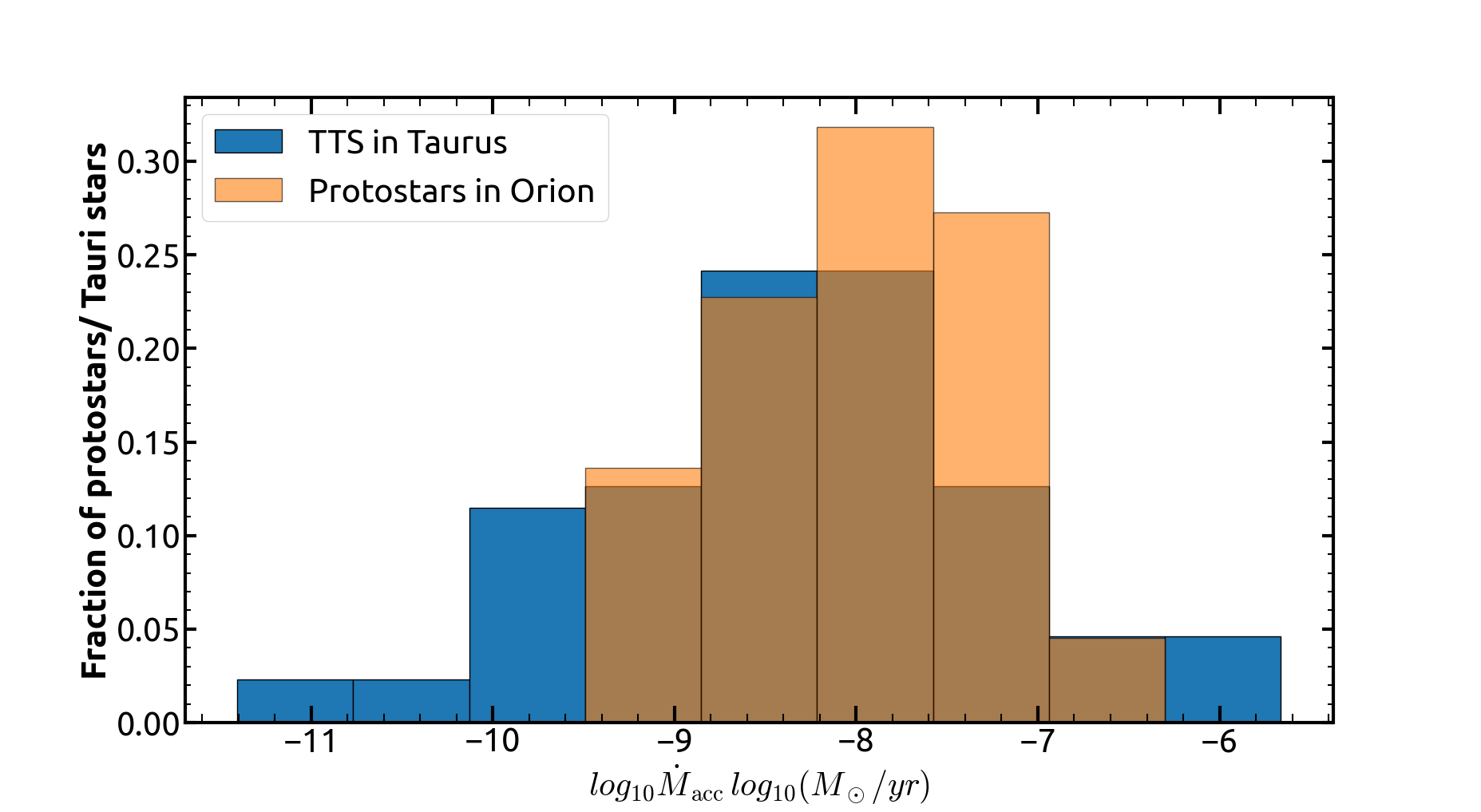}
\caption{The distribution of mass accretion rate of protostars in Orion (orange histogram) and mass accretion rates from well-studied T-Tauri stars from Taurus. }
\label{ch2figmacc}
\end{figure}

\begin{equation} 
    \mathrm{log}_{10}(L_\mathrm{acc}/L_\odot) = (1.13 \pm 0.05)\times \mathrm{log}_{10}(L_{H\alpha}) + (1.74\pm 0.19)
\end{equation}

\noindent
Accretion luminosity is connected to mass accretion rate  ($\dot {M}$) as \citep{Hartmann16}
\begin{equation}
    L_\mathrm{acc}  =  \frac{G\,M_* \dot{M}_\mathrm{acc}}{R_*}  \left( 1 - \frac{R_*}{R_{in}} \right)
\end{equation}
where M$_*$ and R$_*$ are stellar mass and radius and $R_\mathrm{in}$ is the disk truncation radius from which the gas falls onto the star. For T-Tauri stars $R_\mathrm{in}$ is typically assumed to be $\sim$ 5~R$_*$ \citep{gullbring1998}.  Therefore, the above Equation can be written as 
\begin{equation}
    {\dot{M}}_\mathrm{acc} = (1 -  R_* /R_\mathrm{in}) \times (L_\mathrm{acc}R_* / GM_*) 
    \sim 1.25 (L_\mathrm{acc}R_* / GM_*)
\end{equation}

\noindent
We have considered $R_*$/$M_*$ $\sim$ 5 $R_\odot$ /$M_\odot$ (similar to the values used for T-Tauri e.g., \citealt{Kenyon98}, \citealt{2008ApJ...681..594H} ) and used above Equation to estimate ${\dot {M}}_\mathrm{acc}$ for all the protostars.

In Figure \ref{ch2figmacc} (also see Table \ref{Table3}) we show the distribution of the ${\dot {M}}_\mathrm{acc}$ derived for the protostars. We also show the mass accretion rates derived for a well-studied sample of T-Tauri stars from Taurus \citep{Furlan11} using the latest LAMOST DR7 spectra \citep{Luo22}. The median mass accretion rate derived for the protostars is $\sim 1.3\times10^{-8}$ $M_\odot/yr$. The mass accretion rate derived for the T-Tauri sample is $\sim 5 \times 10^{-9}$ $M_\odot/yr$. Thus we find that the mass accretion rates of protostars in our sample and that of T-Tauri stars are similar. Similar values of accretion rates in the range of $1\times10^{-10}$ $M_\odot/yr$  to $1\times10^{-7}$ $M_\odot/yr$ (using optical wavelength) have been derived by \cite{white04, Kenyon98, Riaz15}. At such low (and steady) mass accretion rates, a Sun-like star cannot be formed in $\sim$~0.5 Myr. This strongly suggests that either episodic accretion is the dominant mode of accretion in protostars or that most of the stellar mass is accreted very early on during the Class 0 stage {\citep[e.g.,][]{2005ApJ...633L.137V,2012MNRAS.427.1182S,2022ApJ...924L..23Z,2023pcsf.conf..197M}. }

\begin{table*}[]
\centering
\begin{tabular}{lcccccccc}
Name  & $L_\mathrm{bol}$ & $T_\mathrm{bol}$ & $G_{RP}$ & $A_V$& Instrument/Telescope & Date  & EW H$\alpha$ &   $\mathrm{log}_{10}$ (${\dot {M}}_{acc}$) \\
 & ($L_\odot$)& (K) & & & &  & ($\AA$) &  ($M_\odot/yr$) \\ 
 \\
 \hline
 \\
HOPS 3 & 0.55 & 467&16.1 & 3& ADFOSC/DOT & 2021-02-22   & 122 $\pm$ 6  & -8.25 $\pm$ 0.76 \\
HOPS 45 &8.5 &518&14.3 & 2& ADFOSC/DOT & 2021-02-22 & 184 $\pm$ 11  & -7.23 $\pm$ 0.67\\
HOPS 49 & 0.7&356&14.6&0 & HFOSC/HCT & 2021-12-22 & 23 $\pm$ 1 & -9.29  $\pm$ 0.86 \\
HOPS 58 &4.5 &620&14.5&5 & ADFOSC/DOT & 2021-02-22   & 39 $\pm$ 2 &  -7.53 $\pm$ 0.7 \\
HOPS 59 & 49.5& 528&13.6 &3 & HFOSC/HCT & 2021-12-23 & 40 $\pm$ 1 &   -7.82 $\pm$ 0.72 \\
HOPS 70A & 6.9& 619&14.0&1 & ADFOSC/DOT & 2020-12-18   & 29 $\pm$ 2  & -8.57 $\pm$ 0.79 \\
HOPS 70B & 6.9& 619&14.4&1 & ADFOSC/DOT & 2021-02-22   & 7 $\pm$ 2  & -9.26 $\pm$ 0.86 \\
HOPS 71A &5.6&277 &14.0&0 &ADFOSC/DOT & 2021-02-22   & 24 $\pm$ 1 & -9.03 $\pm$ 0.84\\
HOPS 107 &5&472&15.5&4 & ADFOSC/DOT & 2021-02-22   & 30 $\pm$ 4  & -8.49 $\pm$ 0.78\\
HOPS 134 & 7.8&781&13.6&1 & ADFOSC/DOT & 2021-02-22   & 64 $\pm$ 3 &   -8.13  $\pm$ 0.75 \\
HOPS 134 & 7.8&781&13.6&1 & HFOSC/HCT & 2021-12-23  & 90 $\pm$ 2  & -7.96 $\pm$ 0.73 \\
HOPS 170A &2.5& 832&13.2&0& ADFOSC/DOT & 2021-02-22   & 127 $\pm$ 5  & -7.79 $\pm$ 0.76\\
HOPS 170B &2.5& 832&13.2&0& ADFOSC/DOT & 2021-02-22  & 60 $\pm$ 1  & -8.16 $\pm$ 0.70 \\
HOPS 194 & 12.7& 645&11.8&0& ADFOSC/DOT & 2021-02-22   & 59 $\pm$ 3  & -7.59 $\pm$ 0.73 \\
HOPS 194 & 12.7& 645&11.8&0& HFOSC/HCT & 2019-12-04 &    36 $\pm$ 1 &   -7.84 $\pm$ 0.69 \\
HOPS 194 & 12.7& 645&11.8&0& HFOSC/HCT & 2021-12-23  & 70 $\pm$ 1 &  -7.51 $\pm$ 0.78 \\
HOPS 199 &0.2&576&15.8&3 & ADFOSC/DOT  & 2020-12-18  & 76 $\pm$ 3 & -8.42 $\pm$ 0.66 \\
HOPS 235 & 4.8&680&11.6 &1 & ADFOSC/DOT & 2020-12-18 & 67 $\pm$ 5 & -7.16 $\pm$ 0.66\\
HOPS 235 & 4.8&680 &11.6 &1 & HFOSC/HCT & 2019-12-04 & 65 $\pm$ 2  & -7.17 $\pm$ 0.64 \\
HOPS 235 & 4.8&680 &11.6 &1 & HFOSC/HCT & 2021-12-23  & 105 $\pm$ 2 & -6.94 $\pm$ 0.67\\
HOPS 260 & 1.7&600&15.6&2& ADFOSC/DOT & 2021-02-22 & 53 $\pm$ 2  & -8.81 $\pm$ 0.81\\
HOPS 294 & 2.8&607&14.1&5& ADFOSC/DOT & 2020-12-18  & 39 $\pm$ 1  & -7.23$\pm$ 0.74 \\
\\
\hline
\\
\end{tabular}%
\caption{{The properties ($L_\mathrm{bol}$, $T_\mathrm{bol}$ and $A_V$) and the Gaia $G_{RP}$ magnitude of the protostars along with equivalent width of H$\alpha$ (EW H$\alpha$) detected in the optical spectra. The mass accretion rate ${\dot{M}}_{acc}$ derived using the EW H$\alpha$ are also listed}.}
\label{Table3}
\end{table*}

\section{Summary}
This work presents our results from an optical search of protostars in the Orion Molecular Cloud (OMC) region with $Gaia$ DR3. Out of the 330 protostars in the HOPS catalog, we detected optical counterparts for 62 sources, thereby increasing the number of optically detected protostars by a factor of 2-3. 

We further show that the optically detected protostars are more evolved than the parent sample and prominently consist of Class I and flat-spectrum sources. The optically detected protostars have an \lbol~ distribution similar to that of the larger parent sample in Orion (listed in \citealt{Furlan16}). The \tbol~ distribution of the $Gaia$ detected protostars, however, is different from the parent sample: the optically detected protostars have a higher \tbol~ when compared to the parent sample. A higher \tbol~ indicates these sources are more evolved.

We obtained optical spectra of the $Gaia$ detected protostars using the Hanle Faint Object Spectrograph Camera (HFOSC) on the Himalayan Chandra Telescope (HCT) and the Aries-Devasthal Faint Object Spectrograph \& Camera (ADFOSC) on the 3.6-m Devasthal Optical Telescope (DOT). For 17 out of 62 optically detected protostars, we were able to obtain low-resolution optical spectra. Furthermore, for 4 out of the 17 protostars, photospheric features such as the TiO bands are detected in the spectra, suggesting that the photosphere can form early on in the star formation process. In these cases, the central (proto)stars appear to have a photosphere similar to that of a late M-type star.

We detect strong emission lines, which we used to compute the mass accretion rates onto these protostars and find them to range between $5\times10^{-10}$ to $1\times10^{-7}$ $M_\odot/yr$, which are similar to those found for pre-main sequence T-Tauri stars. At such low mass accretion rates, a Sun-like star cannot be formed in $\sim$~0.5 Myr. This strongly suggests that either episodic accretion is the dominant mode of accretion in protostars or that most of the stellar mass is accreted very early on during the Class 0 stage.   

\section*{Acknowledgements}
We acknowledge the support of the Department of Atomic Energy, Government of India, under Project Identification No. RTI4002.
This work presents results from the European Space Agency (ESA) space mission, \textit{Gaia}. \textit{Gaia} data are being processed by the \textit{Gaia} Data Processing and Analysis Consortium (DPAC). Funding for the DPAC is provided by national institutions, particularly the institutions participating in the \textit{Gaia} MultiLateral Agreement (MLA). This research has also used NASA's Astrophysics Data System Abstract Service and the SIMBAD database, operated at CDS, Strasbourg, France. This works is based on observations obtained at the 3.6m Devasthal Optical Telescope (DOT), which is a National Facility run and managed
by Aryabhatta Research Institute of Observational
Sciences (ARIES), an autonomous Institute under
Department of Science and Technology, Government
of India.  We would also to thank the staff at IAO, Hanle, and its remote control station at CREST, Hosakote, for their help during the observation runs.  

\section*{Appendix}
\section*{HAPILI}

A major challenge in the upcoming years is going to be the sheer volume of data that the various astronomical facilities will collect. In preparation for this, we have started the development of data reduction pipelines for the Faint Object Spectrograph (FOS) class of instruments. We have developed a data reduction pipeline for Hanle Faint Object Spectrograph Camera  (HFOSC) on the Himalayan Chandra Telescope (HCT) and Aries-Devasthal Faint Object Spectrograph $\&$ Camera (ADFOSC) on the 3.6-m Devasthal Optical Telescope (DOT).   

Since both these instruments are Faint Object Spectrographs, the data reduction steps between the two are very similar. Hence a pipeline capable of reducing the data from both instruments was possible. The python-based pipeline runs PYRAF \citep{2012ascl.soft07011S} in the backend. PYRAF is the python wrapper of IRAF \citep{1986SPIE..627..733T}. PYRAF enables IRAF commands to be called and executed within the python environment.

The pipeline first reads the header of all the fits files and, based on the information in the header, classifies the fits files as bias, flats, objects, and lamp. In the case of HCT, the pipeline can distinguish between the FeNe and the FeAr lamp as well. Next, using standard PYRAF routines, the median combined bias and flat files are generated. The science objects are both bias-corrected as well as flat-fielded. The pipeline next has the option of median combining the science object before or after the data reduction. The pipeline next automatically does aperture identification by running the standard IRAF task "\textit{apall}". Standard parameters such as the gain of the instrument and readout noise have already been coded in the pipeline but can be easily modified. This makes the pipeline versatile enough to be used with other FOS class of instruments as well. The aperture is then automatically traced, with the user having the option to adjust the fit. All of this is carried out automatically in an IRAF terminal. Currently, the only step that requires heavy user interaction is wavelength identification which is done using the IRAF task "\textit{identify}". We are working towards automating this step as well. Next, we used the IRAF task "\textit{refspec}" and "\textit{dispcor}" to the wavelength calibrated spectrum and produced the final spectrum.

\vspace{-1em}



\end{document}